# Valley Vortex States in Sonic Crystals

Jiuyang Lu,[1] Chunyin Qiu,[1*] Manzhu Ke,[1] and Zhengyou Liu[1,2*]

[1]Key Laboratory of Artificial Micro- and Nano-structures of Ministry of Education and School of Physics and Technology, Wuhan University, Wuhan 430072, China

[2]Institute for Advanced Studies, Wuhan University, Wuhan 430072, China

**Abstract.** Valleytronics is quickly emerging as an exciting field in fundamental and applied research. In this Letter, we study the acoustic version of valley states in sonic crystals and reveal a vortex nature of such states. Besides the selection rules established for exciting valley polarized states, a mimicked spin Hall effect of sound is proposed further. The extraordinary chirality of valley vortex states, detectable in experiments, may open new possibility in sound manipulations, which is appealing to scalar acoustics since there is no spin degree of freedom inherently. Besides, the valley selection enables a handy way to create vortex matter in acoustics, in which the vortex chirality can be controlled flexibly. Potential applications can be anticipated with the exotic interaction of acoustic vortices with matter, such as to trigger the rotation of the trapped microparticles without contact.

In condensed matter physics, it is always of great interest to explore new internal quantum degrees of freedom of electrons. Recently, the descrete valley degree of freedom, also viewed as pseudospin, is attracting rapid growing attention [1-8]. Valley pseudospin, which labels the degenerate energy extrema in momentum space, has been widely observed in conventional semiconductors and trendy two-dimensional crystals (e.g. graphene and $MoS_2$). The intervalley scattering occurs scarcely due to the large separation in momentum, which makes the valley a good index to characterize the electron states. Like the spin in spintronics, the valley index is poentially a new carrier of information and thus useful in modern electronic devices, leading to the concept of valleytronics [1-4]. Recently, much effort has been devoted to generate and detect the valley polarized current [9-11]. Many exciting phenomena associated with valley contrasting properties have been theoretically predicted and experimentally observed, such as valley filters [2,5,6] and valley hall effects [3,4,7,8].

Because of the similarity of linear waves in periodical structures, valley-like frequency dispersions may also exist in the artificial crystals for classical waves, such as in photonic crystals and sonic crystals (SCs). In these systems, where both the coupling strength among meta-atoms and the symmetry of crystal can be flexibly tailored [12], the involved phenomena can be readily observed in their macroscopic characteristics. In fact, analog to the novel quantum phenomena associated with the conic dispersions, many intriguing wave transport properties have been demonstrated in the artificial crystals, such as *Zitterbewegung* oscillations [13], extremal transmissions [14], and extinction of coherent back-scatterings [15]. Besides, the one-way edge modes that inherently stem from the topological property of crystal states have also been extensively observed in the optic [16] and acoustic systems [17,18].

In this Letter, the concept of valley states is introduced to SCs for acoustic waves and the vortex nature of such states is revealed clearly. Besides the usual momentum matching mechanism, an azimuthal rule for selectively exciting one of the degenerated valley states is established based on the rotational symmetry of crystal. In contrast with atomic crystals where the population and detection of purely polarized valley states often resort to external fields (e.g. strain [9], magnetic [10], and polarized light [11]), the acoustic valley states can be directly excited by external sound stimuli, and detected from the field distributions inside and outside the SCs. The distinguishable valley signature (i.e., vortex chirality) brings us a new degree of freedom to manipulate sound. The excitation of acoustic valley states also enables us to produce a compact array of acoustic vortices with controllable chirality, which is unattainable through conventional approaches based on transducer arrays. Similar to the valley electrons carrying orbital magnetic moments [3], the acoustic valley vortex states carry orbital angular momenta as well. This is particularly meaningful in the acoustic system that inherently lacks the angular momentum generated by spin polarizations. Considering the interaction with matter, besides the implications in fundamental studies, the vortex array created by crystal states may open up new application avenues, such as patterning and rotating microparticles without contact.

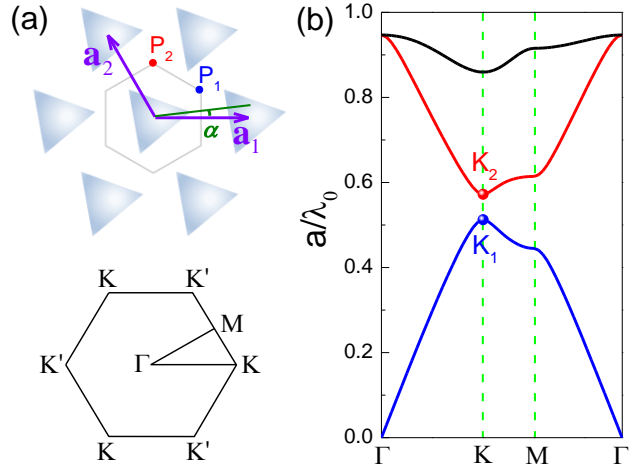

FIG. 1. (color online) (a) Schematics of a hexagonal SC made of regular triangle steel rods immersed in water, where $P_1$ and $P_2$ indicate the positions with $C_3$ symmetry. Bottom panel: The corresponding FBZ. (b) The band structure of the SC with angle $\alpha = 10^\circ$.

As depicted in Fig. 1(a), the SC consists of a hexagonal array of regular triangle steel rods immersed in water [19], where the lattice constant $|\mathbf{a}_1|=|\mathbf{a}_2|=a$, the rotation angle $\alpha = 10^\circ$, and the filling ratio of rod $0.24$. Figure 1(b) gives the dispersion relation of this SC along several typical directions. It displays a pair of well-defined extrema (i.e. valley states) in each corner of the first Brillouin zone (FBZ), which are separated by an *omnidirectional* band gap from the dimensionless frequency $a/\lambda_0 = 0.51$ to $0.57$, with $\lambda_0$ being the wavelength in water. In contrast with the gapless graphene structure, the band gap of this SC stems from the breaking of spatial inversion symmetry [12]. Below the properties of the valley states at K-point (denoted by $K_1$ and $K_2$ for the lower and higher frequency ones) are focused and those for the inequivalent $K'$-point can be deduced directly from the time-reversal (TR) symmetry.

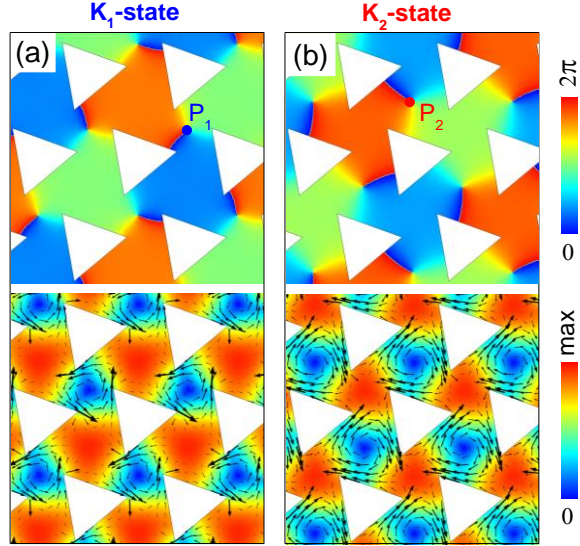

FIG. 2. (color online) Pressure field distributions for the valley states $K_1$ (a) and $K_2$ (b), where the top and bottom panels display (by color) the phase and amplitude patterns, respectively. The additional arrows in the bottom panels indicate the corresponding time-averaged Poynting vectors. Here the sound field in steel rods (white region) is not provided due to the weak penetration.

Similar to electronic valley states, the acoustic valley states also exhibit exotic chirality. This can be observed from the pressure fields in Fig. 2, where the top and bottom panels display (by color) the phase and amplitude distributions, $\phi(\mathbf{r})$ and $|p(\mathbf{r})|$. In the positions of high symmetry (i.e., threefold rotational symmetry $C_3$), $P_1$ and $P_2$, the pressure amplitudes vanish and the phases become singular for $K_1$- and $K_2$-states, respectively. The phase singularity can be characterized by a quantized topological charge: $n=+1$ for $K_1$-state and $n=-1$ for $K_2$-state. For those inequivalent states the chirality is reverse due to the TR-symmetry. The above pressure distribution reveals a typical feature of vortex field, as shown more clearly from the arrows in the lower panels of Fig. 2, the spatial distributions of the time-averaged Poynting vectors $\langle \mathbf{S} \rangle = (2\rho_0 \omega)^{-1} |p|^2 \nabla \phi$, where $\rho_0$ is the mass density in water and $\omega$ is the angular frequency of sound field. As stated above, the chirality of valley vortices can be controlled by exciting the valleys of different frequencies or momenta. In fact, the vortex chirality can also be switched by rotating the orientation of steel rods.

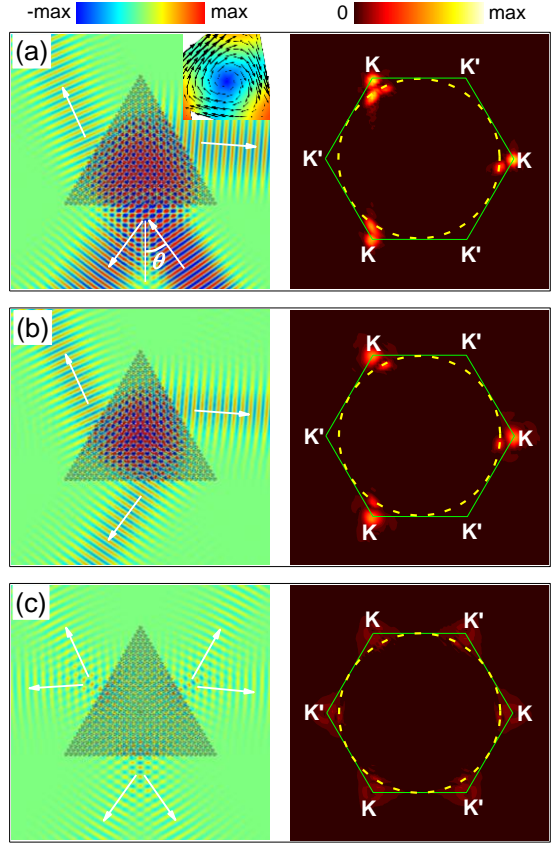

FIG. 3. (color online) Selection rules exemplified by $K_2$-state. (a) The pressure pattern generated by a Gaussian beam illuminating a finite SC obliquely, where the inset enlarges a vortex inside the SC. (b) The pressure pattern stimulated by a chiral source of topological charge $m=-1$ positioned in the center of sample. (c) The same as (b), but for $m=0$. All right panels display the corresponding Fourier spectra in momentum space, where the green solid and yellow dashed lines indicate the hexagonal FBZ and the isofreqency contour of free space, respectively.

Different from the valley electrons, where the selective population (or detection) of one valley is rather less straightforward [9-11], the valley states in SCs can be directly accessed by external sound stimuli. As shown in Fig. 3(a), a Gaussian beam is impinged upon a regular triangle SC with surface normal orientated along the ΓM direction, where the incident angle is chosen to satisfy the momentum matching with $K_2$-state. By this incidence the sound field inside the SC is well excited, featured with clockwise rotated vortices (see inset, in accordance with $K_2$-state). The suppression of $K'_2$-state can be observed more clearly in the corresponding Fourier-spectrum, in contrast with the bright spots at K-points (the bright spots located on the circle stand for the input and outgoing beams traveling in free space). Note that the intervalley scattering is efficiently avoided by using the specific SC shape plus its surface orientation: at each SC boundary, the $K_2$-state is partly reflected back to itself, and partly refracted out according to the conservation of momentum parallel to the crystal interface. Particularly, the beams leaked out (see white arrows) are experimentally detectable and in turn serve as a good evidence of the valley selection.

The valley states can also be selectively excited by a point-like source with proper chirality [20]. The coupling between the valley state $p_n(\mathbf{r})$ and a chiral source $p_m(\mathbf{r}) = A(r)e^{im\varphi}$ can be described by the integration $C_{n,m} = \iint p_n^* p_m d\mathbf{r}$, where $n$ and $m$ correspond to the topological charges of the valley state and chiral source. Under the $2\pi/3$ rotation transformation $\hat{O}$, the crystal state and the chiral source acquire extra phases, i.e. $\hat{O}p_n = p_n e^{i2n\pi/3}$ and $\hat{O}p_m = p_m e^{i2m\pi/3}$. According to the rotational invariance of the system, $\hat{O}C_{n,m} = C_{n,m} e^{i2(m-n)\pi/3} = C_{n,m}$, an azimuthal selection rule can be derived straightforwardly, i.e. $m-n=3l$, with $l$ being an arbitrary integer. The validity of the selection rule can be demonstrated in Fig. 3(b), the pressure distribution and the corresponding Fourier spectrum stimulated by a chiral source of $m=-1$ positioned at one of the equivalent P$_2$-points (i.e., the locations of the vortex cores in K$_2$- and $K'_2$-states). It is observed that the $K_2$-state is successfully excited due to $m-n=0$, whereas $K'_2$-state is completely suppressed since $m-n=-2$. For comparison, similar data for the case of $m=0$ is provided in Fig. 3(c), which manifests that none of the valley states is excited due to the failure of azimuthal phase matching. (Although weak excitation of the crystal states can be seen from the faint speckles in the Fourier spectrum, a careful observation states that the weakest excitation occurs exactly at the FBZ-corners.)

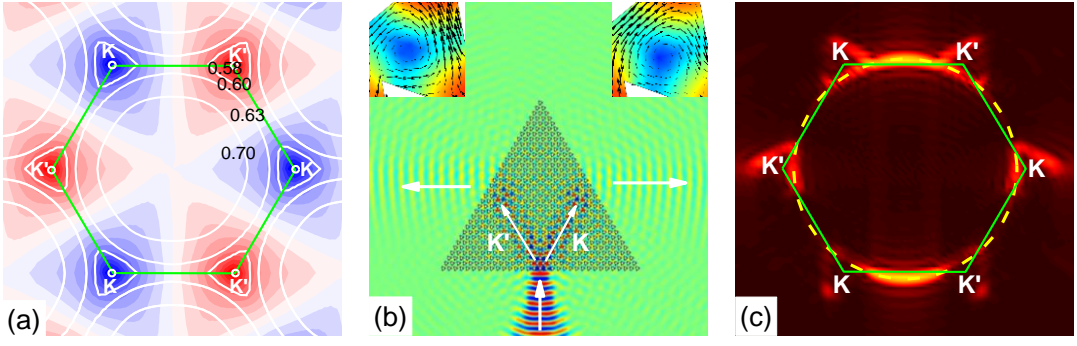

FIG. 4. (color online) (a) The *Q*-factor (color) of the vortex states in the second band, where blue and red indicate the clockwise and anticlockwise rotated vortices, together with several isofrequency contours (white lines) labeled in the momentum space. (b) The pressure distribution stimulated by a Gaussian beam from the bottom, where the insets amplify the anticlockwise and clockwise vortices extracted from the left- and right-going beams inside the SC. (c) The Fourier spectrum of (b).

The vortex feature is not restricted in the corner states, but robustly exists in the states surrounding them. This is confirmed by the field distributions for the crystal states deviated from the valley frequencies, in which the vortex cores could be drifted away from the locations of $C_3$-symmetry. To characterize the quality of the vortex state, a quantity $Q = \int_{Cell} \nabla \times \langle \mathbf{S} \rangle dA$ can be defined. Physically, this quantity is proportional to the acoustic angular momentum localized in each unit cell, $J_z = \int_{Cell} \langle \mathbf{u} \times \rho_0 \dot{\mathbf{u}} \rangle_z dA$, a continuous version of the phonon's pseudospin angular momentum in atomic crystals [21], where $\mathbf{u}$ is the displacement of vibration. As an example, in Fig. 4(a) we present the *Q*-factor for the eigenstates of the second band. It is observed that the quality of the vortex state is optimized at the FBZ-corners and degrades gradually toward the M-point. Therefore, the vortex nature remains rather well even if the frequency deviates from $K_2$-state, e.g., at the dimensionless frequency 0.60, in which notable trigonal warping effect [22] emerges in the isofrequency contour, see Fig. 4(a). It is of interest that such nontrivial distortion of isofrequency contour enables a spatial separation of vortex states carrying opposite chirality, as shown in Fig. 4(b), the pressure distribution stimulated by a narrow Gaussian beam. This chirality-locked beam-splitting phenomenon can be understood from the corresponding Fourier spectrum in Fig. 4(c), where the bright speckles locating on the circle represent the incident, reflected, and refracted beams in free space, and the bright straight bars correspond exactly to the vortex states excited inside the SC. The excitation of the vortex states stems from the broad momentum distribution of the incident beam, and the propagating directions of the split beams can be deduced from the gradient direction of the trigonal isofrequency contour. In some sense, similar to the spin Hall effect of light [23] and surface plasmons [24], this beam-splitting behavior can be viewed as the spin Hall effect of sound, where the valley pseudospin mimics the spin of light. This fascinating observation is particularly meaningful in the longitudianl wave systems without intrinsic spin polarizations, which could stimulate the first experimental realization of the spin Hall effect in acoustics and enable us a novel way to manipulate sound.

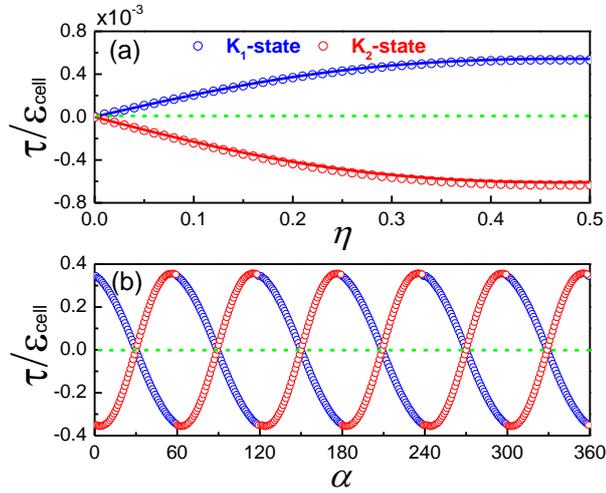

FIG. 5. (color online) (a) The torque imposing on a rubber cylinder with different dissipation coefficient $\eta$, where the circles are evaluated from the *anisotropic* valley vortex states, and the solid lines correspond to the theoretical predictions from the *isotropic* vortices in free space. (b) The torque perceived by the triangle steel scatterer for the SC with different rotation angle $\alpha$.

The above study states that a great number of vortices can be generated simultaneously: the chirality could be unique in the whole SC or opposite in different spatial regions. Comparing with the conventional acoustic vortices (each made of a complex array of piezoelectric transducers or loudspeakers with carefully designed phase lags [25-27]), the vortices created by valley states are compact and easy to fabricate. As a natural character of all chiral phased wave fields, the acoustic valley vortices carry orbital angular momenta and enable a wide range of applications, e.g., to induce mechnical torques on the trapped objects by transferring the angular momenta to matter.

To exhibit the capability of rotating objects by the vortex states in SCs, a dissipative rubber cylinder is placed at the vortex core and the acoustically-induced torque (AIT) $\tau$ is evaluated by integrating the angular momentum density tensor over a circular contour enclosing the cylinder [28]. (Note that the probe cylinder has been verified to be well-trapped in the vortex core, due to the large gradient of the field intensity.) In Fig. 5(a) we present the dimensionless AITs (scaled by the energy density integrated over a unit cell $\varepsilon_{cell}$) calculated for different dissipation coefficient $\eta$ of the probe cylinder [19]. It is observed that, for both $K_1$- and $K_2$-states, the amplitudes of AITs grow linearly with small $\eta$ and saturate gradually at large $\eta$, associated with the signs of AITs consistent with the chirality of vortex states (thus indicating another route to detect the chirality). Recently, it has been theoretically proposed [29-31] and later experimentally validated [25-27] that, for an acoustic vortex in homogenous space, the AIT can be linearly connected with the power absorbed by the object, $\tau = n\omega^{-1}P_{abs}$, with $n$ labeling again the quantized topological charge of vortex. Does this conclusion (justified for *isotropic* vortices) still hold for *anisotropic* vortices? The question has not been addressed so far. Here the absorptions for both valley states are evaluated and the theoretical AITs (lines) are presented in Fig. 5(a). The excellent agreement reveals that the torque-absorption relation is indeed irrelevant to the fine structure of vortex profile. Interestingly, the exotic valley vortex states can also produce torque on the anisotropic scatterer itself. In Fig. 5(b) we present the dimensionless AIT perceived by the triangular steel rod, which exhibits great sensitivity on its orientation and gives rich phenomena. In contrast with the positive slope of $K_2$-state that favors the rotation of scatterer, the negative slope of the AIT in $K_1$-state tends to draw the scatterer back to its equilibrium orientation, which enables the SC more stable to resist undesired rotational perturbations.

In conclusion, we have extended the concept of valley states to SCs for acoustic waves. The acoustic valley states, carrying notable feature of vortices, can be selectively accessed according to the vortex chirality. Benefited from the macroscopic nature of SCs, the valley chirality can be detected in experiments directly, which allows us a brand new manner to control sound. Similar study can also be extended to the other artifical crystals for classical waves. In particular, the extension to vectorial wave systems, e.g., nanostructured plasmonic crystals, will enrich the inherent physics because of the additional coupling between the valley chirialty with intrinsic polorizations, leading to inspiring applications from micromoters to communications in two-dimensional integrated circuits.

**Acknowledgements**

The authors thank Zhenyu Zhang and Qian Niu for fruitful discussions. This work is supported by the National Basic Research Program of China (Grant No. 2015CB755500); National Natural Science Foundation of China (Grant Nos. 11174225, 11374233, 11534013, and J1210061).

**References:**
[1] X. Xu, W. Yao, D. Xiao, and T. F. Heinz, Nat. Phys. **10**, 343 (2014).
[2] A. Rycerz, J. Tworzydlo, and C. W. J. Beenakker, Nat. Phys. **3**, 172 (2007).
[3] D. Xiao, W. Yao, and Q. Niu, Phys. Rev. Lett. **99**, 236809 (2007).
[4] W. Yao, D. Xiao, and Q. Niu, Phys. Rev. B **77**, 235406 (2008).
[5] Z. Wu, F. Zhai, F. M. Peeters, H. Xu, and K. Chang, Phys. Rev. Lett. **106**, 176802 (2011).
[6] J. L. Garcia-Pomar, A. Cortija, and M. Nieto-Vesperinas, Phys. Rev. Lett. **100**, 236801 (2008).
[7] D. Xiao, G. Liu, W. Feng, X. Xu, and W. Yao, Phys. Rev. Lett. **108**, 196802 (2012).
[8] K. F. Mak, K. L. McGill, J. Park, and P. L. McEuen, Science **344**, 1489 (2014).
[9] O. Gunawan et al., Phys. Rev. Lett. **97**, 186404 (2006); K. Takashina, Y. Ono, A. Fujiwara, Y. Takahashi, and Y. Hirayama, Phys. Rev. Lett. **96**, 236801 (2006).
[10] Y. P. Shkolnikov, E. P. De Poortere, E. Tutuc, and M. Shayegan, Phys. Rev. Lett. **89**, 226805 (2002); N. C. Bishop et al., Phys. Rev. Lett. **98**, 266404 (2007); Z. Zhu, A. Collaudin, B.Fauque, W. Kang, and K. Behnia, Nat. Phys. **8**, 89 (2012); D. MacNeill et al., Phys. Rev. Lett. **114**, 037401 (2015).
[11] K. F. Mak, K. He, J. Shan, and T. F. Heinz, Nat. Nanotechnol. **7**, 494 (2012); H. Zeng, J. Dai, W. Yao, D. Xiao, and X. Cui, Nat. Nanotechnol. **7**, 490 (2012); T. Cao et

al., Nat. Commun. **3**, 887 (2012).

[12] J. Lu, C. Qiu, S. Xu, Y. Ye, M. Ke, and Z. Liu, Phys. Rev. B **89**, 134302 (2014).

[13] X. Zhang, Phys. Rev. Lett. **100**, 113903 (2008); Q. Liang, Y. Yan, and J. Dong, Opt. Lett. **36**, 2513 (2011).

[14] R. A. Sepkhanov, J. Nilsson, and C. W. J. Beenakker, Phys. Rev. B **78**, 045122 (2008); X. Zhang, and Z. Liu, Phys. Rev. Lett. **101**, 264303 (2008); S. R. Zandbergen and M. J. A. de Dood, Phys. Rev. Lett. **104**, 043903 (2010); S. Bittner, B. Dietz, M. Miski-Oglu, P. Oria Iriarte, A. Richter, and F. Schafer, Phys. Rev. B **82**, 014301 (2010).

[15] R. A. Sepkhanov, A. Ossipov, and C. W. J. Beenakker, EPL **85**, 14005 (2009); G. Weick, C. Woollacott, W. L. Barnes, O. Hess, and E. Mariani, Phys. Rev. Lett. **110**, 106801 (2013).

[16] F. D. M. Haldane and S. Raghu, Phys. Rev. Lett. **100**, 013904 (2008); Z. Wang, Y. Chong, J. D. Joannopoulos, and M. Soljačić, Phys. Rev. Lett. **100**, 013905 (2008); Z. Wang, Y. Chong, J. D. Joannopoulos, and M. Soljačić, Nature **461**, 772 (2009); Y. Poo, R. Wu, Z. Lin, Y. Yang, and C. T. Chan, Phys. Rev. Lett. **106**, 093903 (2011); M. C. Rechtsman et al., Nature **496**, 196 (2013); A. B. Khanikaev et al., Nat. Mater. **12**, 233 (2013); W. Chen et al., Nat. Commun. **5**, 5782 (2014); L. Lu, J. D. Joannopoulos, and M. Soljačić, Nat. Photonics **8**, 821 (2014); T. Ma, A. B. Khanikaev, S. H. Mousavi, and G. Shvets, Phys. Rev. Lett. **114**, 127401 (2015); L. Wu, and X. Hu, Phys. Rev. Lett. **114**, 223901 (2015).

[17] X. Ni et al., New J. Phys. **17**, 053016 (2015); Z. Yang et al., Phys. Rev. Lett. **114**, 114301 (2015).

[18] P. Wang, L. Lu, and K. Bertoldi, Phys. Rev. Lett. **115**, 104302 (2015); M. Xiao et al., Nat. Phys. **11**, 240 (2015); M. Xiao et al., Nat. Phys. In press; V. Peano, C. Brendel, M. Schmidt, and F. Marquardt, Phys. Rev. X **5**, 031011 (2015).

[19] Material parameters: $\rho = 1.0 g/cm^3$ and $c_l = 1.49 km/s$ for water; $\rho = 7.67 g/cm^3$, $c_l = 6.01 km/s$ and $c_t = 3.23 km/s$ for steel; $\rho = 1.2 g/cm^3$, $c_l = 2.3 km/s$ and $c_t = 0.94 km/s$ for rubber. Here $\rho$, $c_l$, and $c_t$ represent the density, longitudinal velocity, and transverse velocity, respectively. The absorption in rubber is introduced by an imaginary part of the longitudinal velocity $\eta c_l$.

[20] Practically, the point-like chiral source can be realized by a subwavelength circular array of identical point-sources with proper phase lags.

[21] L. Zhang, and Q. Niu, Phys. Rev. Lett. **112**, 085503 (2014); *ibid.* **115**, 115502 (2015).

[22] J. L. Garcia-Pomar, A. Cortijo, and M. Nieto-Vesperinas, Phys. Rev. Lett. **100**, 236801 (2008).

[23] M. Onoda, S. Murakami, and N. Nagaosa, Phys. Rev. Lett. **93**, 083901 (2004); O. Hosten, and P. G. Kwiat, Science **319**, 787 (2008); Q. Guo, W. Gao, J. Chen, Y. Liu, and S. Zhang, Phys. Rev. Lett. **115**, 067402 (2015).

[24] A. A. High et al., Nature **522**, 192 (2015).

[25] K. Volke-Sepulveda, A. O. Santillan, and R. R. Boullosa, Phys. Rev. Lett. **100**, 024302 (2008).

[26] C. E. M. Demore, Z. Y. Yang, A. Volovick, S. Cochran, M. P. MacDonald, and G. C. Spalding, Phys. Rev. Lett. **108**, 194301 (2012).

[27] A. Anhauser, R. Wunenburger, and E. Brasselet, Phys. Rev. Lett. **109**, 034301 (2012).

[28] Here the calculation is carried out from the eigenstate. The radius of the rubber cylinder is $0.1a$, which is small enough and leads to negligible influence on the eigenstate. The result has also been checked in a finite SC by external excitation, which exhibits a weak dependence on the location of vortex core because that once the valley state is excited the vortex pattern in each lattice site keeps almost invariant except the amplitude.

[29] J. Lekner, J. Acoust. Soc. Am. **120**, 3475 (2006).

[30] K. D. Skeldon, C. Wilson, M. Edgar, and M. J. Padgett, New J. Phys. **10**, 013018 (2008).

[31] L. K. Zhang, and P. L. Marston, Phys. Rev. E **84**, 065601 (R) (2011).